# Tuning Interfacial Water Friction through Moiré Twist

*Chenxing Liang[ab], N. R. Aluru[ab]\**

a. Walker Department of Mechanical Engineering, The University of Texas at Austin, Texas 78712, USA

b. Oden Institute for Computational Engineering and Sciences, The University of Texas at Austin, Texas 78712, USA

\* Correspondence to aluru@utexas.edu (N.R.A.)

**ABSTRACT**

Nanofluidics is pivotal in fundamental research and diverse applications, from water desalination to energy harvesting and biological analysis. Dynamically manipulating nanofluidic properties, such as diffusion and friction, presents an avenue for advancement in this field. Twisted bilayer graphene, particularly at the magic angle, has garnered attention for its unconventional superconductivity and correlated insulator behavior due to strong electronic correlations. However, the impact of the electronic properties of moiré patterns in twisted bilayer graphene on structural and dynamic properties of water remains largely unexplored. Computational challenges, stemming from simulating large unit cells using density functional theory, have hindered progress. This study addresses this gap by investigating water behavior on twisted bilayer graphene, employing a deep neural network potential (DP) model trained with a dataset from *ab initio* molecular dynamics simulations. It is found that as the twisted angle approaches the magic angle, interfacial water friction increases, leading to reduced water diffusion. Notably, the analysis shows that at smaller twisted angles with larger moiré patterns, water is more likely to reside in AA stacking regions than AB (or BA) stacking regions, a distinction that diminishes with smaller moiré patterns. This exploration illustrates the potential for leveraging the distinctive properties of twisted bilayer graphene to effectively control and optimize nanofluidic behavior.

# INTRODUCTION

Nanofluidics, a field with broad research appeal, has been extensively studied for its fundamental principles and practical applications. In particular, the behavior of water at the nanoscale when confined within carbon nanotubes [1,2,3] or graphene nanochannels [4,5,6] has been extensively investigated. This exploration has extended to real-world applications such as water desalination, [7,8] energy harvesting, [9,10] biological analysis, [11,12] and the development of nanofluidic memristors. [13,14] The influence of material properties on nanofluidic behavior is a focal point of these investigations. For instance, the wetting properties of materials significantly affect water transport, with hydrophobic carbon nanotubes displaying rapid water transport. [15,16] Additionally, the mechanical properties of materials, such as phonon vibrations, has been observed to enhance fluid transport through phonon-fluid coupling within nano porous membranes. [8] Surface roughness, another material property, has also been shown to influence water contact angles and slip lengths on top of 2-D material such as hexagonal boron nitride (hBN). [17]

Nonetheless, while the influence of material properties such as wetting, mechanical, and surface roughness on nanofluidic behavior has been investigated extensively, the role of electronic properties remains relatively underexplored. Insights from *ab initio* molecular dynamical study by Angelos et al. [18] emphasize that the distinct electronic properties of boron nitride contribute to 3 times higher water friction compared to a graphene surface. Further research indicates a notable discrepancy in water friction between carbon nanotubes and boron nitride nanotubes, attributed to the specific hydrogen–nitrogen interactions between boron nitride and water. [19] Moreover, it has been reported that water encounters quantum friction at the solid-liquid interface due to the coupling between charge fluctuations in the liquid and electronic excitations in the solid. [20]

Consequently, comprehending the coupling between material electronic properties and nanofluidics and leveraging the electronic properties of materials to modulate nanofluidic behavior emerges as a pivotal area for exploration.

Lately, the magic angle twisted bilayer graphene (MATBG) has garnered attention owing to its correlated insulator [21] and unconventional superconductivity behaviors [22] arising from strong electronic correlations. Various moiré patterns, which arise from periodic modulation of lattice mismatch, exist in twisted bilayer graphene with different twisted angles. When the twisted angle approaches the magic angle (approximately 1.08°), coulomb interactions between electrons supersede kinetic energy of electrons, [21] leading to the emergence of topologically nontrivial flat bands and highly localized fermions in real space. Intriguingly, in the MATBG, fermions exhibit localization primarily within the AA stacking region rather than the AB (or BA) stacking region. [21] What adds further intrigue to the twisted bilayer graphene is the tunability of fermion localization, which is represented by reduced fermi velocity, achievable by altering the twist angle. As the angle is adjusted within the range from 5° to the magic angle (1.08°), the fermi velocity decreases and the degree of fermion localization intensifies. [23] However, this phenomenon of localization dissipates when the twist angle exceeds 5° or diverges significantly from the magic angle.

Hence, the exploration of the relationship between the electronic properties of twisted bilayer graphene and nanofluidics presents a compelling area for investigation. Notably, a recent study by Majumder et al. [24] has delved into the influence of the twist angle of bilayer graphene on the structural and dynamic properties of water. Employing classical molecular dynamics simulations, the study observed that while the water distribution structure remained relatively unaffected by changes in the twisted angle, significant sensitivity emerged in dynamical properties like water friction and slip length. Yet, this analysis focused on twisted angles beyond the critical range between the magic angle and 5°, thus overlooking the pivotal role of the electronic properties intrinsic to twisted bilayer graphene. Moreover, the study exhibited disparate trends in water friction with different twist angles when utilizing various water models, highlighting the substantial sensitivity of the friction to the utilized water models within classical molecular dynamics simulations. Additionally, the fixed charges on all atoms and rigid water in classical MD simulations omitted consideration of the dynamic coupling between twisted bilayer graphene and nanofluidics.

These limitations highlight the necessity of investigating the coupling between twisted bilayer graphene (encompassing twist angles from the magic angle to 5°) and nanofluidics through simulations with quantum accuracy. Such an approach can circumvent issues related to the sensitivity of classical water models and can accurately portray the dynamic coupling between twisted bilayer graphene and nanofluidics simultaneously. However, it is important to note that the unit cell of twisted bilayer graphene becomes exceedingly large, with over 10,000 atoms, [23] when the twisted angle approaches the magic angle. Conducting density functional theory calculations under such conditions demands extensive computational resources. Detailed information regarding the unit cell sizes for different twisted bilayer graphene configurations in this study can be found in the supplementary Table S2.

The advent of machine learning has revolutionized scientific computing by addressing complex challenges. Recently, a study on the phonon properties of magic angle twisted bilayer graphene employed a deep neural network interatomic potential. [25] This approach involved training the model using *ab initio* molecular dynamics data of twisted bilayer graphene with smaller unit cells. Once a reliable interatomic potential was established, simulations were conducted on magic angle twisted bilayer graphene. In alignment with this methodology, our research investigates water behavior atop twisted bilayer graphene, encompassing twisted angles ranging from the magic angle to 5°. The approach involves several key steps: Firstly, we generated an *ab initio* molecular dynamics (AIMD) training dataset by running simulations of water atop twisted bilayer graphene with smaller unit cells across twisted angles from 5° to 32°. Subsequently, an interatomic potential based on a deep neural network was trained, followed by deep potential (DP) model [26] validation and testing. This trained model was then applied to study the structural and dynamical properties of water on twisted bilayer graphene within the range of twist angles from the magic angle to 5°.

Our study delves into an analysis of the impact of twisted angle variations, which is related to electronic properties of twisted bilayer graphene, on the structural and dynamical properties of water. Moreover, we provide an explanation of the underlying physics governing the coupling between nanofluidics and the twisted angle bilayer graphene. Finally, we offer a conclusive summary of our findings in this study.

**METHODS**

**Geometry Modeling:** The system schematic, presented in Figure 1 (a), delineates water atop twisted bilayer graphene. The modeling of the twisted bilayer graphene system utilizes the theoretical derivations from Sato et al.'s work. [27] The specific twisted angle is controlled by the index m and n, detailed in Uchida et al.'s work. [23] The resulting unit cell of moiré patterns initially takes a hexagonal form, later converted to an orthogonal unit cell for analytical convenience. The overall system, inclusive of water positioned on top of twisted bilayer graphene, is created using Packmol. [28] In the training dataset for *ab initio* molecular dynamics (AIMD) simulations, various systems featuring water atop twisted bilayer graphene are investigated across eight different twisted angles (6.01°, 7.34°, 9.43°, 13.17°, 16.43°, 21.79°, 28.80°, and 32.20°) with smaller unit cells. Detailed geometric specifications, inclusive of the sizes, counts of water molecules and carbon atoms, and the trajectory length of AIMD simulations, are presented in Supplementary Table S1. In Deep Potential molecular dynamic (DPMD) simulations, employing a well-trained deep neural network interatomic potential, systems comprising water on top of twisted bilayer graphene are investigated across six different twisted angles (1.08°, 1.89°, 2.88°, 3.89°, 4.41°, and 5.09°) with larger unit cells. Specifics regarding geometric parameters, quantities of water molecules and carbon atoms, as well as the trajectory length of DPMD simulations, are detailed in Supplementary Table S2.

***Ab initio* Molecular Dynamics:** The *ab initio* molecular dynamics (AIMD) simulations were conducted using the CP2K software package (version 2022.1) [29] based on the density functional theory. [30] To accurately capture the structural and dynamical properties of water, the SCAN (Strongly Constrained and Appropriately Normed) exchange-correlation functional [31] and the Goedecker−Teter−Hutter (GTH) pseudopotentials [32] were specifically employed. The energy cutoff was set at 800 Ry for these simulations. The wave function was optimized using the orbital transformation (OT) method, with a convergence criterion of $1\times10^{-6}$ atomic units. A time step of 0.5 femtoseconds was utilized, maintaining a constant temperature of 330 K throughout the simulations. For the initialization of the AIMD simulations, the initial structure was obtained from classical molecular dynamics (MD) simulations performed using the LAMMPS software.[33] These classical MD simulations utilized the SPC/E water model [34] and force fields developed by Wu et al.[35] Following a 1 nanosecond equilibration process, ensuring a well-equilibrated initial structure for AIMD, the structure is relaxed using density functional theory to prepare it as initial configuration for AIMD simulations.

**DP Training:** The deep neural network potential (DP) model training process utilized the smoothed version of the DeePMD model within the DeepMD-kit package.[26] The model decomposes the total energy of the atomic system into atomic contributions:

$$E(R) = \sum_{i=1}^{N} E_i \qquad (1)$$

where $E_i$ represents the atomic energy, reliant upon the local atomic environment. A distance cutoff of 8 Å was set to define the local atomic configuration, with a smoothing initiation point at 1.0 Å. This deep neural network comprises an embedding network and a fitting network. The embedding network transforms local coordinates $R_i$ to the input descriptor, preserving the translational, rotational, and permutational symmetries of the atomic system. In our training, the embedding network encompasses three hidden layers with 50, 100, and 200 nodes, respectively. The fitting network uses the descriptor as input, producing atomic energy $E_i$ and atomic force $f_i$ via three hidden layers, each with 240 nodes. The model parameters are obtained by minimizing the loss function: $L = \frac{p_e}{N}\Delta E^2 + \frac{p_f}{3N}\Delta F_i^2$. Where, $p_e$ and $p_f$ are prefactors, while $\Delta E, \Delta F_i$ denote the root mean square errors of the predicted energy and force. During training, $p_e$ ranges from 0.02 to 1, while $p_f$ decreases from 1000 to 8. The Adam optimizer with an initial learning rate of 1e-3, decaying to 3.51e-8, was used for training over 5,000,000 steps on NVIDIA A100 GPUs to minimize model error.

In Figure 2 (a), the training process of DP model is depicted, showcasing the training based on AIMD simulation data from eight systems, each involving water positioned atop twisted bilayer graphene with varying angles from 6.01° to 32.20° (6.01°, 7.34°, 9.43°, 13.17°, 16.43°, 21.79°, 28.80° and 32.20°) with smaller unit cells. Once a reliable DP model is established, DPMD simulations with quantum accuracy can be undertaken to explore systems featuring smaller twisted angles with larger unit cells. The training process included active learning techniques aimed at enhancing the breadth and quality of the training dataset, extending the coverage across the phase space. This strategy aligns with a similar concept highlighted in Luana et al.'s work. [36] Following the completion of training for the DPMD model, simulations were performed on systems featuring random twisted angles among 6.01°, 7.34°, 9.43°, 13.17°, 16.43°, 21.79°, 28.80° and 32.20°. After 50 ps of DPMD simulations, the system structure was utilized as the new starting point for AIMD simulations. These AIMD simulations were then incorporated into the training dataset to further refine the DPMD model, ensuring its adaptability and precision.

**DP model Validation and Test:** In Figure 2 (b), we conducted validation and testing processes for the trained DP model by excluding the training dataset with twisted angles of 6.01°, 13.17°, and 27.80°. Using the same training approach outlined in the DP Training section, we derived a DP model. This DP model was then employed to conduct DPMD simulations on systems comprising water atop twisted bilayer graphene with these excluded angles.

Subsequently, we compared the trajectories from DPMD simulations with those from AIMD simulations for validation and testing. For this, we evaluated various parameters such as bond length, power spectra, velocity autocorrelation function, mean square displacement, hydrogen bonds number, and water density profiles. These comparisons were performed between DPMD simulations with twisted angles of 13.17° and 27.80° and their corresponding AIMD simulations for validation. Additionally, for testing purposes, we compared the results obtained from DPMD simulations to those from AIMD simulations for the case with a twisted angle of 6.01°.

The comparison figures illustrating the DPMD and AIMD results for both validation and testing are included in Supplementary Figures S1, S2, and S3. These figures demonstrate a close match between the DPMD and AIMD results. This alignment between DPMD and AIMD results signifies the reliability and accuracy of the DP model in capturing the nanofluidic behavior across varied twisted angles. In prior research by Liu et al., [25] findings also demonstrate the DPMD model's efficacy in accurately forecasting the properties of magic angle twisted bilayer graphene when the DPMD model was trained using AIMD simulation data derived from smaller unit cells. These results affirm the DPMD model's capability to successfully predict the properties of magic angle twisted bilayer graphene.

**DPMD Simulation:** To reduce the computational expenses associated with obtaining quantum-accurate data through *ab initio* calculations, we conducted DPMD simulations, leveraging a well-established deep neural network interatomic potential. The systems under study consisted of water and twisted bilayer graphene characterized by six distinct twisted angles, spanning from the magic angle to 5° (1.08°, 1.89°, 2.88°, 3.89°, 4.41°, and 5.09°). The unit cell configuration for each twisted bilayer graphene is detailed in Figure 1(b)-(g). The simulations operated at a temperature of 330 K and a timestep of 0.5 fs, extending over a duration of 100 ps. For the initial configuration of the DPMD simulation, the system was first equilibrated using the SPC/E water model [34] and Wu et al.'s force field [35] within the LAMMPS platform. Additionally, the DPMD simulations for each system were executed across five different ensembles with distinct initial velocities to ensure robust statistical representation.

**RESULTS & DISCUSSION**

In this section, we present a summary of the findings acquired through our DPMD simulations. Furthermore, we delve into an in-depth discussion concerning the fundamental principles underlying the coupling between the twisted bilayer graphene and the dynamics of nanofluidics.

**Water density profile**

The water density profile, representing the water distribution along normal direction of the surface, is extensively analyzed to depict the overall distribution characteristics. Utilizing the trajectory data from DPMD simulations, the water density profile atop twisted bilayer graphene with varying twisted angles (1.08°, 1.89°, 2.88°, 3.89°, 4.41°, 5.09°) is compared in Figure 3 (a). Interestingly, the observations illustrate an identical water density profile despite the alteration in the twisted angle of the graphene layers. This uniformity across different twisted angles demonstrates that the water distribution along the surface's perpendicular direction remains incentive to changes in the twisted angle.

**Friction Coefficient and Diffusion coefficient**

To further investigate the dynamic properties of water on twisted bilayer graphene with varying twisted angles, an analysis was performed to determine the friction coefficient and diffusion coefficient. In each equilibrium DPMD simulation, the friction coefficient was computed using the Green–Kubo formula,[37] entailing the time integral of the force autocorrelation function defined as:

$$\lambda_{GK}(\tau) = \frac{1}{Ak_BT}\int_0^\tau dt \langle F(0)F(t) \rangle \quad (1)$$

where, $A$ signifies the interfacial lateral area, $\langle ... \rangle$ denotes an ensemble average, and $F(t)$ represents the lateral force applied by graphene onto the water at time $t$. This force, $F(t)$, is assessed as the cumulative force acting on all water molecules within a specific configuration, averaged across both in-plane dimensions (x, y),[38] and recorded at each time-step (0.5 fs). The friction coefficient of the system is ideally determined at the extended time limit:

$$\lambda = \lim_{\tau \to \infty} \lambda_{GK}(\tau) \quad (2)$$

However, as time progresses, the integral in Equation (1) diminishes to zero due to the system's finite lateral extent.[39] Therefore, a common practice is to determine $\lambda_{GK}(\tau)$ by observing a plateau.[18,40] Research has indicated that accurately retrieving friction relies on a distinct timescale separation between the decay time and the memory time of the force autocorrelation function when taking the maximum of $\lambda_{GK}(\tau)$.[41] Given that we're investigating interface behaviors where such timescale separation is absent, we approximate λ as the friction coefficient at correlation times of 1 ps.[38] Additionally, we conducted 5 separate ensembles of DPMD simulations for each case to enhance the statistical robustness. The error bars for water friction in each case were determined based on these five different ensembles. The collective result of the average water friction atop twisted bilayer graphene, spanning from the magic angle to 5°, is depicted in Figure 2 (b) and detailed in Table 1. The conversion of the friction coefficient unit from the metal unit in LAMMPS to the reported unit ($\times 10^4$ N • s/m³) is elucidated in Supplementary Equation S1. Remarkably, our findings reveal an increased water friction as the twisted angle approaches closer to the magic angle. Specifically, the water friction coefficient at the magic angle, where fermions are localized, is approximately 75% higher than at 5°, where fermion localization diminishes. This discrepancy indicates the sensitivity of water friction to the twisted angle. Furthermore, these results underscore the correlation between the dynamic properties of water and the electronic properties exhibited by twisted bilayer graphene.

To provide a comprehensive understanding of water's dynamic properties, we extended our analysis to include the calculation of the diffusion coefficient for the first layer of water near the graphene interface. This approach was selected due to the potential inaccuracies associated with computing the diffusion of the entire water box atop graphene, especially when dealing with a limited number of water layers. The rapid diffusion of the water at the water-vapor interface can impact the diffusion of the overall water box, potentially resulting in an inflated diffusion coefficient compared to bulk water. For a detailed examination of the diffusion among different water layers atop twisted bilayer graphene, reflecting the influence from the water-vapor interface, refer to Supplementary Figure S4. The accelerated diffusion at the water-vapor interface can obscure the impact of the twisted bilayer graphene layer on water diffusion. However, the water layer proximate to the graphene interface remains unaffected by the water-vapor interface diffusion. To validate this, we investigated the diffusion of the first water layer adjacent to the graphene interface under varying numbers of water layers. Observations show that the diffusion of the water layer near the interface exhibits little sensitivity to the number of water layers present. For a schematic of the system and detailed results, please refer to Supplementary Figure S5 and Figure S6. The mean-squared displacement (MSD) of oxygen atoms of water in the first layer near the interface is first obtained, and the diffusion coefficient of water near the interface can be subsequently calculated using the Einstein relation: [42]

$$D = \lim_{t \to \infty} \frac{1}{2dt} \langle [r(t) - r(0)]^2 \rangle \tag{3}$$

In the equation, $d$ denotes the system's dimension, and $r(t)$ signifies the coordinate of the oxygen atom at time $t$. By averaging five different ensembles of the DPMD trajectory, we obtained the mean-squared displacement (MSD) plot. The average diffusion coefficients for each case are depicted in Figure 2 (c) and outlined in Table 1, accompanied by the attached statistical error for the calculated diffusion coefficients.

According to our calculations, the diffusion coefficient of water at the magic angle is approximately 27% lower than that at 5°. Notably, as the twisted angle diminishes and nears the magic angle, the diffusion coefficient of water decreases. This trend in water diffusion coefficient aligns well with the observed water friction results. In summary, the outcomes from water friction and water diffusion analyses indicate that as the twisted angle decreases and nears the magic angle, water friction intensifies, and the diffusion coefficient decreases.

**Water Probability Distribution Analysis**

The investigation into the fundamental physics behind the relationship between twisted angle and water friction and diffusion coefficients involved a detailed analysis of water's behavior. In a study by Gabriele et al., [18] the spatial distribution probability of water on surfaces such as hexagonal boron nitride exhibited a tendency for water to accumulate in specific regions, creating a more corrugated energy surface and leading to increased water friction. Similarly, we conducted a comparative analysis of the water spatial distribution relative probability on top of two twisted bilayer graphene with different twisted angles. With the most significant disparities found between the cases at the magic angle and 5°, we calculated the water distribution relative probability for these two scenarios. Notably, for magic angle twisted bilayer graphene, our findings (depicted in Figure 4 (a)) revealed a higher water distribution probability in the AA stacking region compared to the AB (or BA) stacking region within a single unit cell. Additionally, in the supercell, we observed a similar localization effect for water (depicted in Figure 4 (b)), echoing the behavior seen in localized fermions. [21] This suggests an interaction between the localized fermions in the AA stacking region and the water molecules. Conversely, in the case of twisted bilayer graphene with a 5° twisted angle, water distribution probability was relatively uniform (as shown in Figure 4 (c) and (d)). This analysis shows that the heightened water friction on magic angle twisted bilayer graphene could arise from the increased water distribution probability in the AA stacking region, implying a coupling between the localized fermions in this region and the nearby water molecules. On the contrary, the lower water friction observed on twisted bilayer graphene at a 5° angle can be attributed to the more uniform distribution of water.

To delve deeper into the interplay between localized fermions and water molecules, we conducted an analysis of water orientation distribution and interaction energies within both the AA and AB (or BA) stacking regions. The water dipole orientation is calculated and its distribution over these regions is depicted in Figure 4 (e). The horizontal axis of the graph represents the cosine values of the angle between the water dipole direction and the normal vector of the twisted bilayer graphene surface, while the y-axis illustrates the associated probabilities. Our findings revealed notable distinctions in the water dipole distribution when situated atop the AA stacking region as opposed to the AB (or BA) stacking region. Particularly, a divergence is evident when the angle between the water dipole direction and the surface normal exceeds 90° (indicated by a cosine smaller than 0), denoting a distinct orientation wherein the water molecule's two hydrogen atoms are in proximity to the graphene surface. Analysis of water orientation probabilities demonstrates a preference for the placement of hydrogen atoms near the graphene surface within the AA stacking region, indicating a prevailing '2-leg' geometry. Conversely, the AB (or BA) stacking region exhibits a tendency towards a 'flat lying' geometry, with the oxygen atom closer to the surface. The schematic of the '2-leg' geometry and 'flat lying' geometry can be found in Figure 4(e). This variance in water molecule behavior mirrors findings from Thiemann et al.'s [19] research on water transport in hexagonal boron nitride nanotubes and carbon nanotubes, where oxygen confers slipperiness and hydrogen induces stickiness. To further comprehend the distinct behavior of water on various stacking regions, we quantified the interaction energy between water molecules and both the AA and AB (or BA) stacking regions. As illustrated in Table 2, the interaction energy is higher between water and the AA stacking region compared to the AB (or BA) stacking region. The interaction energy is determined by contrasting the energy difference between a system featuring water on magic angle twisted bilayer graphene and another system involving water and magic angle twisted bilayer graphene positioned at a distance without interaction. Further details regarding the estimation of the interaction energy can be found in Supplementary Figure S7 and Equation S2.

Instead of being distinct for magic angle twisted bilayer graphene, the AA stacking and AB (or BA) stacking regions are also observed in the moiré patterns of twisted bilayer graphene when considering larger twisted angles, such as 5.09°. This prompts the question: what causes the variation in water friction and diffusion coefficients with differing twisted angles? Notably, we have observed a significant alteration in the sizes of the AA and AB (or BA) stacking regions as we vary the twisted angles. As the twisted angle approaches the magic angle, the size of the AA stacking region notably expands. Supplementary Table S1 provides detailed information on the unit cell sizes within the moiré pattern. Our assumption is that the differences in friction arise from the varying sizes of the distinct Moiré patterns. At the magic angle, where larger AA stacking regions are prevalent, a substantial number of water molecules tend to occupy these regions for more extended periods, encountering minimal disturbance from water molecules within the AB (or BA) stacking regions due to the substantial distance between these regions. The heightened interaction energy between the AA stacking region and water molecules contributes to increased water friction. Conversely, in scenarios where the twisted angle is larger, leading to smaller distances between AA and AB (or BA) stacking regions, fewer water molecules reside on the AA stacking regions and are more prone to disruption by adjacent water molecules in the AB (or BA) stacking regions. Consequently, the water experiences reduced interaction with the AA stacking region, resulting in decreased water friction.

To validate this assumption, we calculated the water molecule residence time autocorrelation function on top of AA stacking regions within Moiré patterns with twisted angles at 1.08° and 5.09°, respectively. The residence time autocorrelation function, denoted by $\zeta(t)$ and defined as $\zeta(t) = \frac{\langle \beta(t+t_0)\beta(t_0) \rangle}{\langle \beta(t_0)\beta(t_0) \rangle}$, [43] involves $t_0$ as the reference time and $t$ as the elapsed time from $t_0$. Here, $\beta(t + t_0)$ equals 1 if a water molecule exists in the contact layer between $t_0$ and $t + t_0$, otherwise, $\beta(t + t_0)$ equals 0. The $\langle\ \rangle$ denotes the ensemble average. Physically, $\zeta(t)$ represents the probability of a water molecule's residence in the contact layer over time. The resultant calculations are portrayed in Figure 4 (f). The depicted $\zeta(t)$ values show a higher value at the magic angle in comparison to the 5.09° angle, indicating a greater probability of water molecules residing atop the AA stacking region at the magic angle. The residence time of a water molecule in the AA stacking region can be quantified by the correlation time $\tau = \int_0^\infty \zeta(t)dt$. Estimations suggest that the residence time of water molecules on top of AA stacking regions at the magic angle should be notably longer than that at the 5.09° angle. Consequently, we attribute the varying water friction with different twisted angles to the dimensions of the AA and AB (or BA) stacking regions within different Moiré patterns. Water molecules tend to exhibit extended residence times on larger AA stacking regions and experience fewer disturbances when the twisted angle is smaller. This tendency contributes to higher water friction and lower diffusion.

**Conclusion**

This paper explores the structural and dynamic properties of water molecules atop moiré patterns of twisted bilayer graphene, spanning different twisted angles from 1.08° to 5.09°, utilizing a deep neural network potential trained from an *ab initio* Molecular Dynamics (AIMD) dataset. The AIMD dataset encompasses water behavior on twisted bilayer graphene ranging from 6.01° to 32.20°. Through the construction of a reliable Deep Potential (DP) model, accurately describing quantum interactions between water and moiré patterns, molecular dynamics simulations are executed, offering quantum precision at a computational cost equal to classical MD simulations.

Analysis using the DP model reveals that, despite similar water density profiles across distinct moiré patterns, dynamic properties, such as diffusion and friction, are notably sensitive to the twisted angle. Decreasing the twisted angle towards the magic angle results in higher water friction and correspondingly lower water diffusion. Examination of water density distribution probabilities within the graphene plane demonstrates a higher tendency for water to reside in the AA stacking regions as opposed to the AB (or BA) stacking regions in magic angle twisted bilayer graphene, whereas the distribution probability is uniform when the twisted angle is 5.09°. Moreover, this study illustrates differing geometric preference of water molecules atop AA and AB (or BA) stacking in Moiré patterns. Water molecules tend to exhibit a '2-leg' geometry, positioning their hydrogen atoms in proximity to the graphene surface when atop AA stacking regions, while adopting a 'flat-lying' geometry, with the oxygen atom closer to the surface, on AB (or BA) stacking regions. Energy analysis also indicates higher interaction energy between water molecules and the AA stacking region compared to the AB (or BA) stacking region. Conclusively, the variance in friction is attributed to the differing sizes of AA and AB (or BA) stacking regions in various Moiré patterns. Specifically, in the magic angle twisted bilayer graphene, water molecules demonstrate longer residence times and encounter fewer disturbances in the AA stacking region compared to the 5.09° twisted bilayer graphene. This disparity rationalizes the heightened water friction observed atop the magic angle twisted bilayer graphene.

This study exemplifies the adaptability of nanofluidic behavior through the manipulation of electronic properties within twisted bilayer graphene. It elucidates the fundamental relationship between moiré patterns and nanofluidic behavior, potentially facilitating practical applications in water desalination, energy harvesting, and biological analyses by leveraging the electronic properties of materials to control nanofluidic behavior.

## AUTHOR INFORMATION

**Corresponding Author**

* Correspondence to aluru@utexas.edu (N.A)**Author Contributions**

N.A conceived and supervised the project; C. L. collected data, performed analysis, and interpreted the results; C.L. created an initial draft of the manuscript; All authors reviewed the results and approved the final version of the manuscript.

**ACKNOWLEDGEMENT**

The work received support from the Center for Enhanced Nanofluidic Transport (CENT), an Energy Frontier Research Center funded by the U.S. Department of Energy, Office of Science, Basic Energy Sciences under Award No. DE-SC0019112. Computing resources on Frontera and Lonestar6 were made available by the Texas Advanced Computing Center (TACC) at The University of Texas at Austin under Allocation Nos. CHE23031 and DMR22008, respectively. Additionally, the Frontera Computational Science Fellowship provided essential support for computational resources and traveling.

**FIGURES**

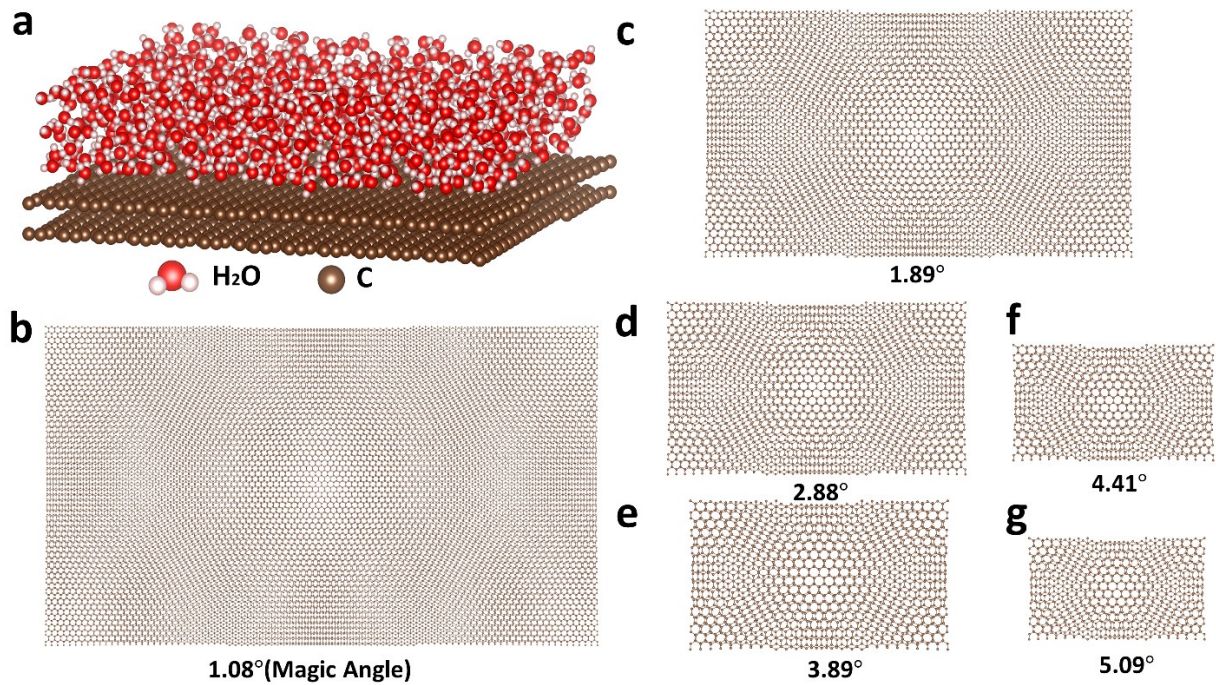

**Figure 1. Schematics of the moiré nanofluidics system and unit cells for various moiré patterns** (a) The interfacial water on top of twisted bilayer graphene. (b) Unit cell of the moiré pattern at 1.08° twisted angle. (c) Unit cell of the moiré pattern at 1.89° twisted angle. (d) Unit cell of the moiré pattern at 2.88° twisted angle. (e) Unit cell of the moiré pattern at 3.89° twisted angle. (f) Unit cell of the moiré pattern at 4.41° twisted angle. (g) Unit cell of the moiré pattern at 5.09° twisted angle.

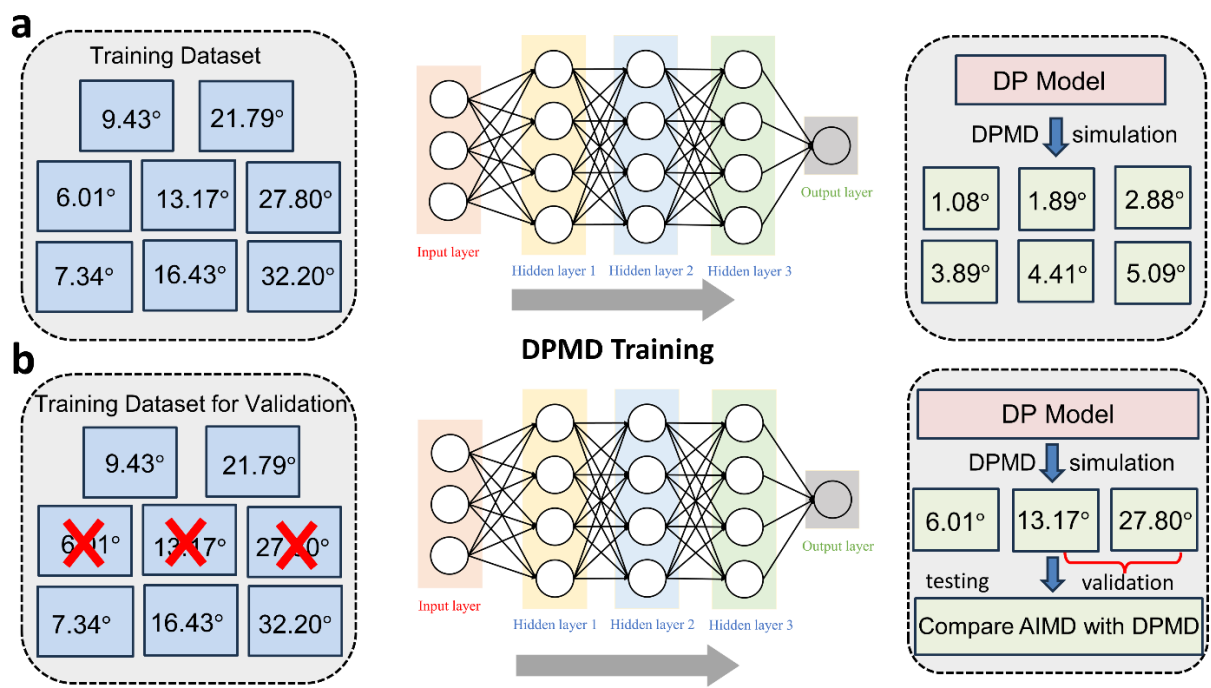

**Figure 2. Machine learning framework utilized for DPMD training, validation, and testing** (a)The training approach, utilizing the AIMD simulation trajectory data for water on twisted bilayer graphene with twisted angles ranging from 5.09° to 32.20°. Subsequently, the well-trained DPMD model is utilized to conduct DPMD simulations for twisted bilayer graphene systems spanning angles from 1.08° to 5.09°. (b) The validation and testing framework for the DPMD model, wherein cases at 6.01°, 13.17°, and 28.80° are omitted from the training data. DPMD simulations are performed for validation at 13.17° and 28.80° cases, while the 6.01° case is used for testing purposes.

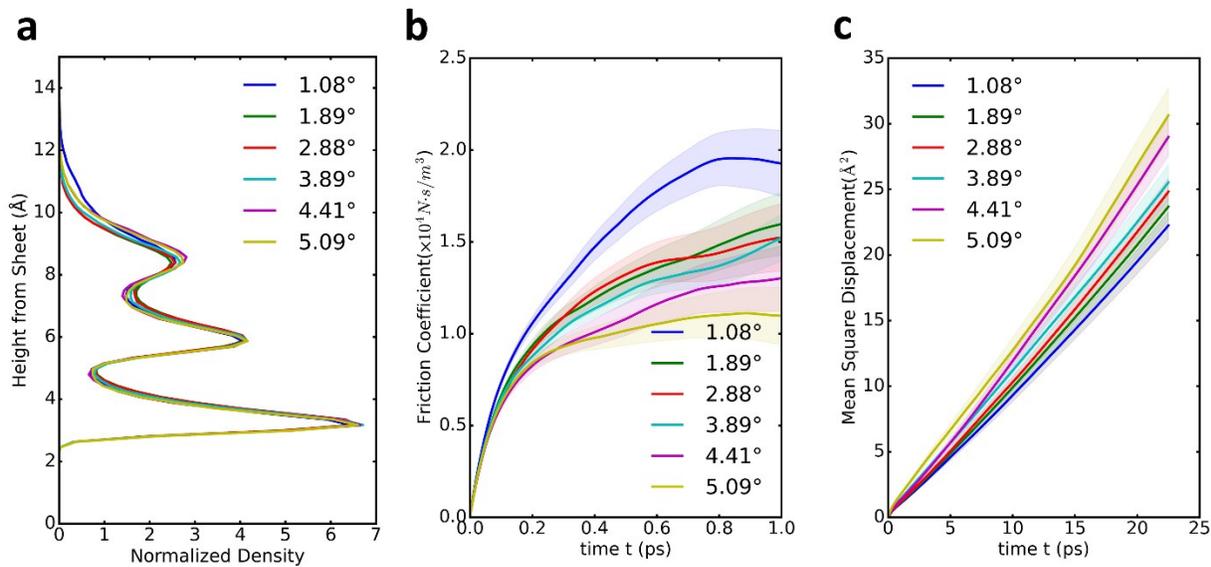

**Figure 3**. **Results obtained from DPMD simulations, focusing on water density profiles, water friction, and water diffusion.** (a) The water density profile perpendicular to the graphene surface, where no distinguishable differences are observed among various cases. (b) Water friction on twisted bilayer graphene with different twisted angles, showing an increase in water friction as the twisted angle decreases. (c) The mean square displacement plot, demonstrating reduced water diffusion as the twisted angle diminishes.

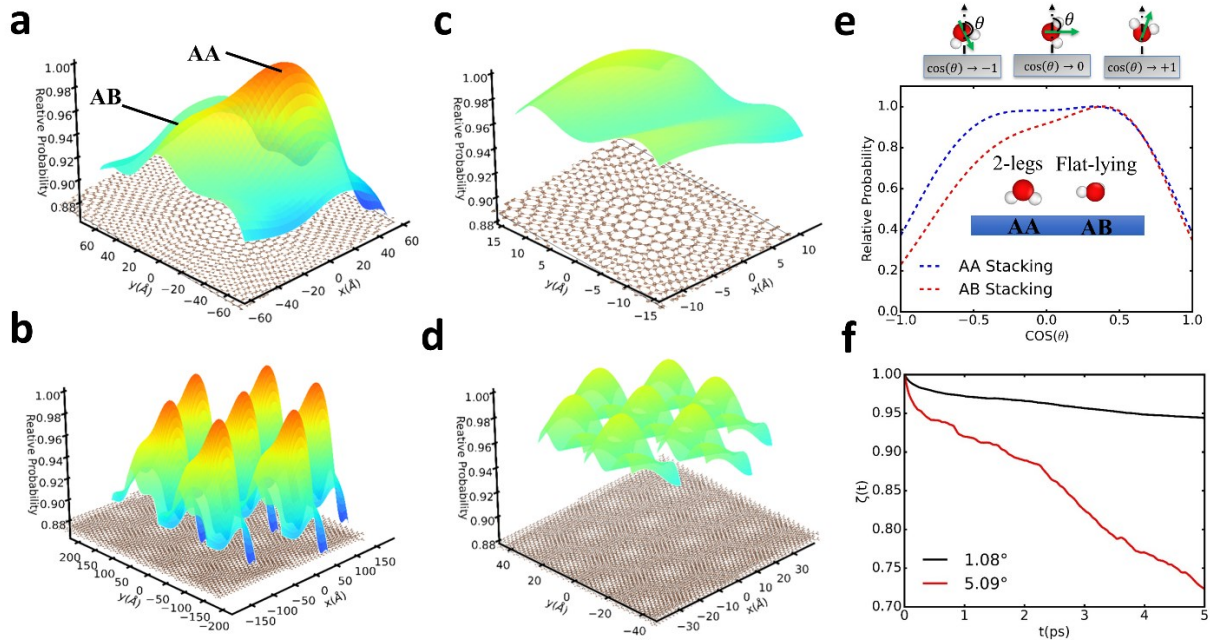

**Figure 4. The physics underlying moiré patterns, encompassing water distribution probability, water orientation, and auto-correlation function of water residence in the AA stacking.** (a) Water distribution probability on a unit cell of the moiré pattern with twisted angle 1.08°. (b) Water distribution probability on a super cell of the twisted bilayer graphene at 1.08°. (c) Water distribution probability on a unit cell of the moiré pattern with twisted angle 5.09°. (d) Water distribution probability on a super cell of twisted bilayer graphene at 5.09°. (e) Water orientation distribution on both AA and AB (or BA) stacking regions of 1.08° twisted angle bilayer graphene. (f) The auto-correlation function of water residence time in the AA stacking region of twisted bilayer graphene at 1.08° and 5.09°.

**Table 1: Averaged friction coefficient and diffusion coefficient of water on top of twisted bilayer graphene with different twisted angles ranging from magic angle to 5°**

| Twisted Angle | Friction Coefficient ($\times 10^4$ N·s/m$^3$) | Diffusion Coefficient ($\times 10^{-8}$ m$^2$/s) |
|---|---|---|
| 1.08° | 1.9267 ± 0.1801 | 0.1648 ± 0.04599 |
| 1.89° | 1.5964 ± 0.1702 | 0.1754 ± 0.04573 |
| 2.88° | 1.5228 ± 0.1844 | 0.1838 ± 0.04603 |
| 3.89° | 1.5204 ± 0.1261 | 0.1891 ± 0.0599 |
| 4.41° | 1.3018 ± 0.1776 | 0.2147 ± 0.06370 |
| 5.09° | 1.0975 ± 0.1561 | 0.2270 ± 0.09311 |

**Table 2: Interaction energy between water and AA/AB (or BA) stacking region of Moiré pattern in magic angle twisted bilayer graphene.**

|  | AA stacking region | AB (or BA) stacking region |
|---|---|---|
| Interaction Energy/per water (eV) | 0.05991 ± 0.004788 | 0.05015 ± 0.005290 |

**REFERENCES**

(1) Liang, C.; Rayabharam, A.; Aluru, N. R. Structural and Dynamical Properties of H2O and D2O under Confinement. *The Journal of Physical Chemistry B* **2023**, *127* (29), 6532-6542. DOI: 10.1021/acs.jpcb.3c02868.
(2) Calegari Andrade, M. F.; Pham, T. A. Probing Confinement Effects on the Infrared Spectra of Water with Deep Potential Molecular Dynamics Simulations. *The Journal of Physical Chemistry Letters* **2023**, *14* (24), 5560-5566. DOI: 10.1021/acs.jpclett.3c01054.
(3) Neklyudov, V.; Freger, V. Water and Ion Transfer to Narrow Carbon Nanotubes: Roles of Exterior and Interior. *The Journal of Physical Chemistry Letters* **2021**, *12* (1), 185-190. DOI: 10.1021/acs.jpclett.0c03093.
(4) Xie, Q.; Alibakhshi, M. A.; Jiao, S.; Xu, Z.; Hempel, M.; Kong, J.; Park, H. G.; Duan, C. Fast water transport in graphene nanofluidic channels. *Nature Nanotechnology* **2018**, *13* (3), 238-245. DOI: 10.1038/s41565-017-0031-9.
(5) Neek-Amal, M.; Lohrasebi, A.; Mousaei, M.; Shayeganfar, F.; Radha, B.; Peeters, F. M. Fast water flow through graphene nanocapillaries: A continuum model approach involving the microscopic structure of confined water. *Applied Physics Letters* **2018**, *113* (8), 083101. DOI: 10.1063/1.5037992 (acccessed 11/6/2023).
(6) Zhao, W.; Qiu, H.; Guo, W. A Deep Neural Network Potential for Water Confined in Graphene Nanocapillaries. *The Journal of Physical Chemistry C* **2022**, *126* (25), 10546-10553. DOI: 10.1021/acs.jpcc.2c02423.
(7) Dong, Z.; Zhang, C.; Peng, H.; Gong, J.; Zhao, Q. Modular design of solar-thermal nanofluidics for advanced desalination membranes. *Journal of Materials Chemistry A* **2020**, *8* (46), 24493-24500, 10.1039/D0TA09471D. DOI: 10.1039/D0TA09471D.
(8) Noh, Y.; Aluru, N. R. Phonon-Fluid Coupling Enhanced Water Desalination in Flexible Two-Dimensional Porous Membranes. *Nano Letters* **2022**, *22* (1), 419-425. DOI: 10.1021/acs.nanolett.1c04155.
(9) Zhang, Z.; Wen, L.; Jiang, L. Nanofluidics for osmotic energy conversion. *Nature Reviews Materials* **2021**, *6* (7), 622-639. DOI: 10.1038/s41578-021-00300-4.
(10) Tong, X.; Liu, S.; Crittenden, J.; Chen, Y. Nanofluidic Membranes to Address the Challenges of Salinity Gradient Power Harvesting. *ACS Nano* **2021**, *15* (4), 5838-5860. DOI: 10.1021/acsnano.0c09513.
(11) Yamamoto, K.; Ota, N.; Tanaka, Y. Nanofluidic Devices and Applications for Biological Analyses. *Analytical Chemistry* **2021**, *93* (1), 332-349. DOI: 10.1021/acs.analchem.0c03868.
(12) Chantipmanee, N.; Xu, Y. Nanofluidics for chemical and biological dynamics in solution at the single molecular level. *TrAC Trends in Analytical Chemistry* **2023**, *158*, 116877. DOI: https://doi.org/10.1016/j.trac.2022.116877.
(13) Sheng, Q.; Xie, Y.; Li, J.; Wang, X.; Xue, J. Transporting an ionic-liquid/water mixture in a conical nanochannel: a nanofluidic memristor. *Chemical Communications* **2017**, *53* (45), 6125-6127, 10.1039/C7CC01047H. DOI: 10.1039/C7CC01047H.
(14) Robin, P.; Emmerich, T.; Ismail, A.; Niguès, A.; You, Y.; Nam, G. H.; Keerthi, A.; Siria, A.; Geim, A. K.; Radha, B.; et al. Long-term memory and synapse-like dynamics in two-dimensional nanofluidic channels. *Science* **2023**, *379* (6628), 161-167. DOI: 10.1126/science.adc9931 (acccessed 2023/11/05).
(15) Majumder, M.; Chopra, N.; Andrews, R.; Hinds, B. J. Enhanced flow in carbon nanotubes. *Nature* **2005**, *438* (7064), 44-44. DOI: 10.1038/438044a.
27

For Table of Contents only

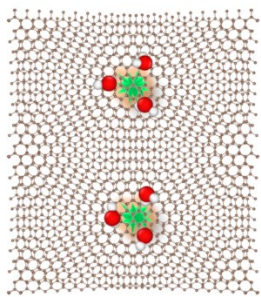 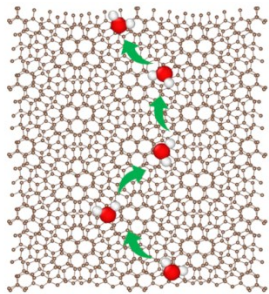

Smaller twisted angle, higher friction    Larger twisted angle, lower friction

# Supporting Information for: Tuning Interfacial Water Friction through Moiré Twist

*Chenxing Liang[ab], N. R. Aluru[ab]\**

a. Walker Department of Mechanical Engineering, The University of Texas at Austin, Texas 78712, USA

b. Oden Institute for Computational Engineering and Sciences, The University of Texas at Austin, Texas 78712, USA

\* Correspondence to aluru@utexas.edu (N.R.A.)

1. **Deep Potential (DP) Model Validation and Testing**

In this section, the comparison of various properties including bond length/angle of water, vibrational spectra of water, mean square displacement (MSD) of water, velocity auto correlation function (VACF) of water and the water density distribution calculated from DP model and AIMD data is conducted to validate the test the DP model.

**1.1 Deep Potential (DP) Model Validation of water on 13.17° twisted bilayer graphene**

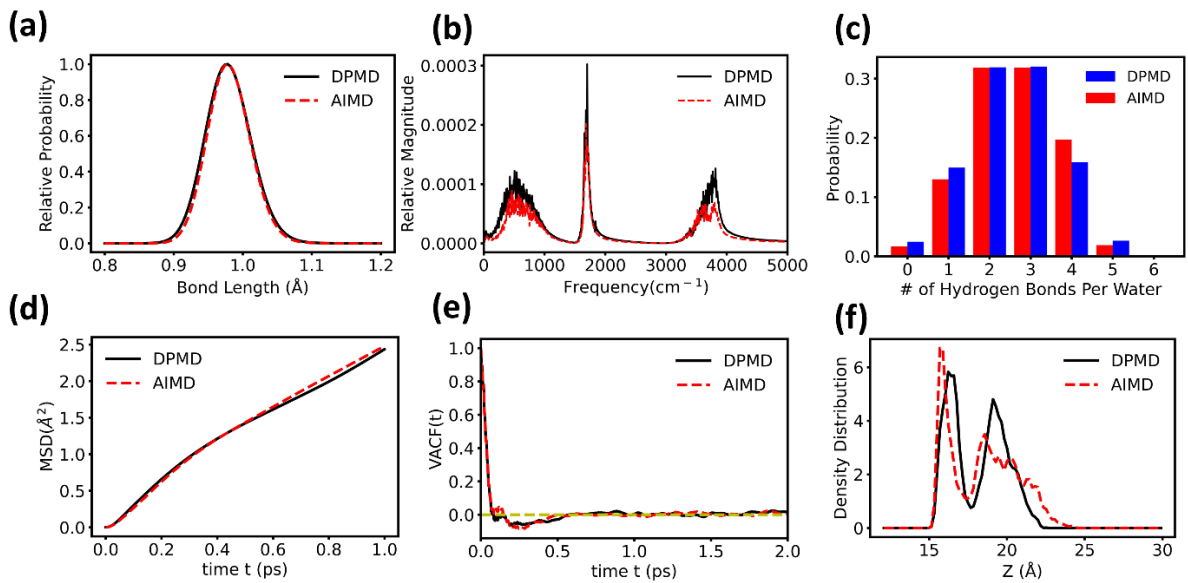

**Figure S1. Comparison between DPMD and AIMD of water on top of 13.17° twisted bilayer graphene.** (a) Bond length of water calculated from DPMD and AIMD. (b) Power spectra of water calculated from DPMD and AIMD. The three peaks in the power spectra represent the libration model, HOH bending model and OH bond stretching model, respectively. (c) Distribution of the number of hydrogen bonds per water calculated from DPMD and AIMD. (d) Mean square displacement (MSD) of water calculated from DPMD and AIMD. (e) Velocity auto-correlation function (VACF) of water calculated from DPMD and AIMD. (f) Water density profile along the normal direction of the graphene surface calculated from DPMD and AIMD.

## 1.2 Deep Potential (DP) Model Validation of water on 27.80° twisted bilayer graphene

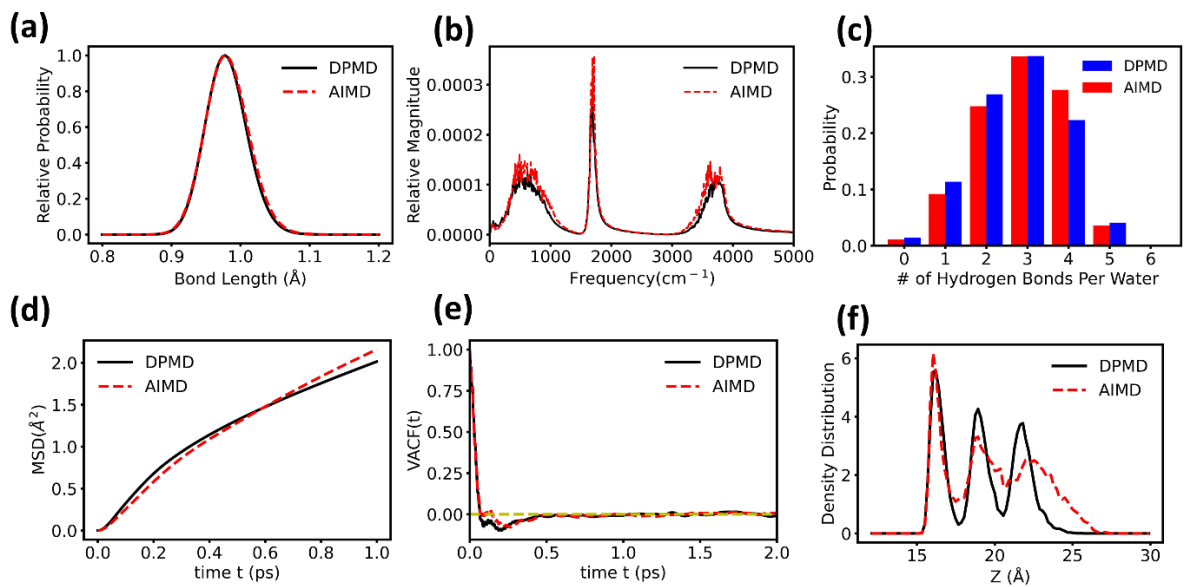

**Figure S2. Comparison between DPMD and AIMD of water on top of 27.80° twisted bilayer graphene.** (a) Bond length of water calculated from DPMD and AIMD. (b) Power spectra of water calculated from DPMD and AIMD. (c) Distribution of the number of hydrogen bonds per water calculated from DPMD and AIMD. (d) Mean square displacement (MSD) of water calculated from DPMD and AIMD. (e) Velocity auto-correlation function (VACF) of water calculated from DPMD and AIMD. (f) Water density profile along the normal direction of the graphene surface calculated from DPMD and AIMD.

### 1.3 Deep Potential (DP) Model Testing of water on 6.01° twisted bilayer graphene

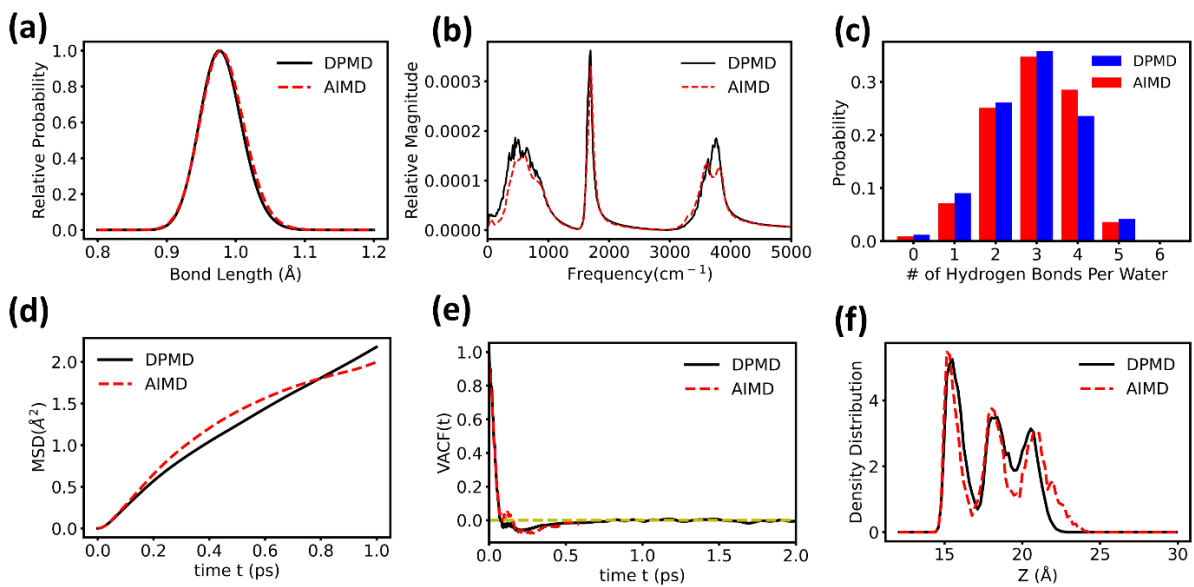

**Figure S3. Comparison between DPMD and AIMD of water on top of 6.01° twisted bilayer graphene.** (a) Bond length of water calculated from DPMD and AIMD. (b) Power spectra of water calculated from DPMD and AIMD. (c) Distribution of the number of hydrogen bonds per water calculated from DPMD and AIMD. (d) Mean square displacement (MSD) of water calculated from DPMD and AIMD. (e) Velocity auto-correlation function (VACF) of water calculated from DPMD and AIMD. (f) Water density profile along the normal direction of the graphene surface calculated from DPMD and AIMD.

2. **Detailed information on parameters of the system (water on top of twisted bilayer graphene with various twisted angles) in training dataset and DPMD simulations.**

Table S1: Summary of the system parameters in training dataset of AIMD simulations

| Twisted Angle | Unit Cell Size $X \times Y \times Z$ (Å) | # of carbon atoms | # of water molecules | AIMD simulation length (fs) |
|---|---|---|---|---|
| 6.01° | 40.75×23.53×33.42 | 728 | 294 | 2,067.0 |
| 7.34° | 33.36×38.52×33.42 | 976 | 300 | 3,534.5 |
| 9.43° | 26.00×30.02×33.42 | 592 | 300 | 5,102.5 |
| 13.17° | 18.60×10.74×33.42 | 152 | 54 | 15342.0 |
| 16.43° | 29.92×17.28×33.42 | 392 | 150 | 2327.5 |
| 21.79° | 22.62×13.06×33.42 | 224 | 150 | 12260.0 |
| 27.80° | 26.70×15.41×33.42 | 312 | 150 | 10525.5 |
| 32.20° | 30.83×17.80×33.42 | 416 | 150 | 8488.0 |

Table S2: Summary of the system parameters in DPMD simulations

| Twisted Angle | Unit Cell Size $X \times Y \times Z$ (Å) | # of carbon atoms | # of water molecules | DPMD simulation length (ps) |
|---|---|---|---|---|
| 1.08° | 225.83×130.3×36.48 | 22328 | 9900 | 100 |
| 1.89° | 129.59×74.82×36.48 | 7352 | 3000 | 100 |
| 2.88° | 85.17×49.18, ×36.48 | 3176 | 1300 | 100 |
| 3.89° | 62.97×36.36×36.48 | 1736 | 730 | 100 |
| 4.41° | 55.52×32.08×36.48 | 1352 | 600 | 100 |
| 5.09° | 48.17×27.81 ×36.48 | 1016 | 450 | 100 |

3. The conversion of the friction coefficient unit from the metal unit in LAMMPS to the reported unit ($\times 10^4$ N·s/m³)

$$\frac{1}{\text{Å}^2 \cdot J} \cdot \left(\frac{eV}{\text{Å}}\right)^2 \cdot ps$$
$$= \frac{1}{10^{-20} \times m^2 \cdot N \cdot m} \cdot \left(\frac{1.60218 \times 10^{-19} \cdot N \cdot m}{10^{-10} \times m}\right)^2 \cdot 10^{-12} \qquad \text{S1}$$
$$= (1.96 \times 10^{-14}) \times 10^4 \; N \cdot s/m^3$$

4. The influence of the water-vapor interface on the diffusion of the whole water box as well as the diffusion of the first layer water near the graphene interface.

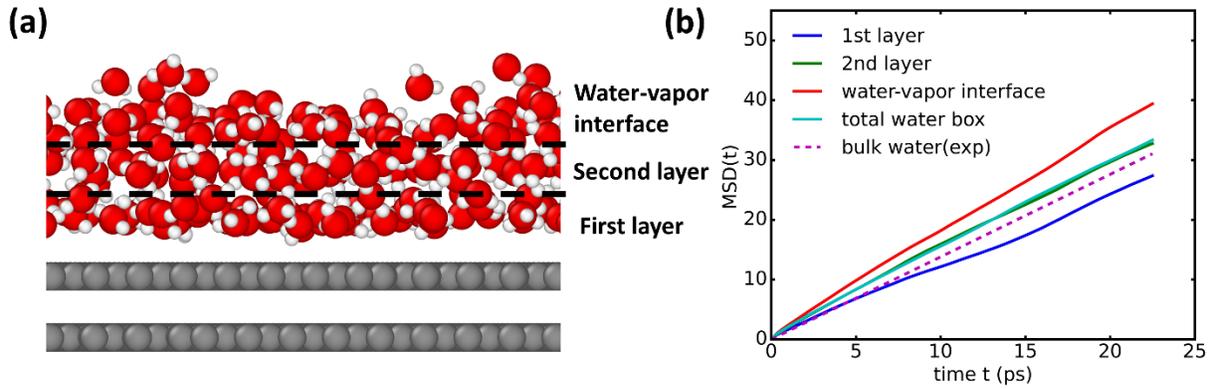

**Figure S4: The diffusion of water molecules at different layers.** (a) The system depicts the first layer, second layer, and the interface with water vapor. (b) Examining the mean square displacement of water across different layers indicates that the rapid diffusion at the water-vapor interface can cause the overall water box diffusion to exceed that of bulk water diffusion. [1] Therefore, calculating the diffusion of the entire water box in this study might not be accurate.

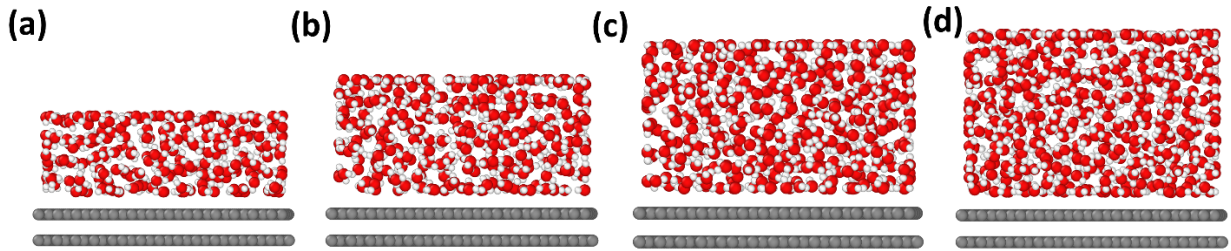

**Figure S5: Simulation systems used to study the influence from the water-vapor interface.** (a) The water box on top of 7.43° twisted bilayer graphene, which consists of 3 layers of water. (b) The water box on top of twisted bilayer graphene, which consists of 4 layers of water. (c) The water box on top of twisted bilayer graphene, which consists of 5 layers of water. (d) The water box on top of twisted bilayer graphene, which consists of 6 layers of water.

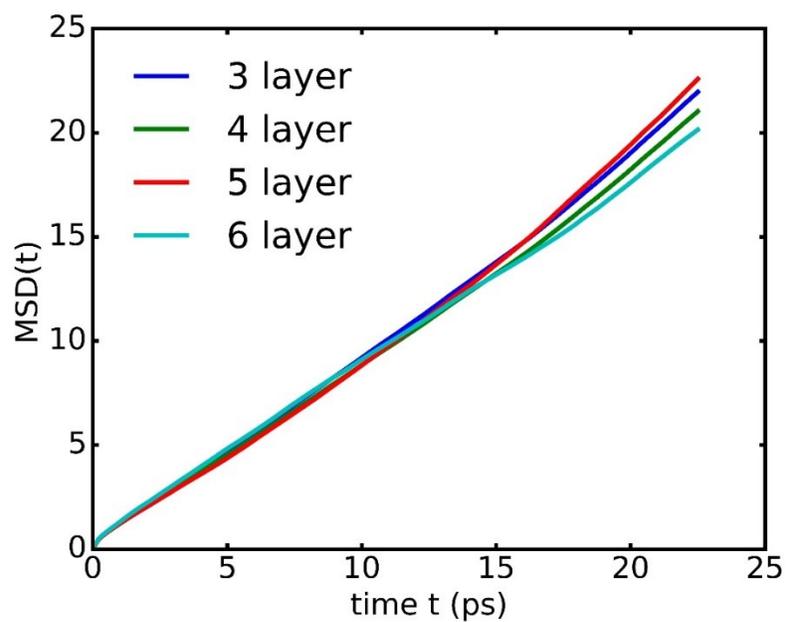

**Figure S6:** The diffusion coefficient of the first layer of water near the graphene interface in various systems which consists of 3 water layers, 4 water layers, 5 water layers and 6 water layers, respectively.

5. **Detailed information on the calculation procedure of interaction energy between water and AA/AB (or BA) stacking region of magic angle twisted bilayer graphene.**

**(a)** System 1 with water-AA/AB stacking interaction

**(b)** System 2 without water-AA/AB stacking interaction

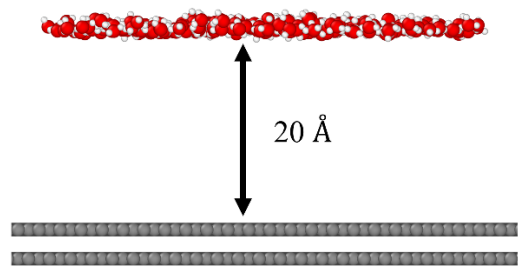

20 Å

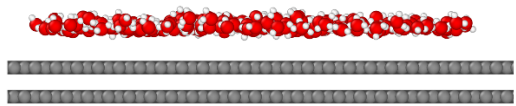

**Figure S7: Two systems that are utilized to calculate the interaction energy between water and AA/AB (or BA) stacking region of magic angle twisted bilayer graphene.** (a) system 1 with water-AA/AB (or BA) stacking interaction. (b) system 2 without water-AA/AB (or BA) stacking interaction. The separation distance between water and the graphene surface is 20 Å.

The interaction energy between water and AA/AB (or BA) stacking region is calculated as:

$$E_{interaction} = \frac{(E_2 - E_1)}{\# \; of \; water \; molecules} \qquad \text{S2}$$

Which is the energy difference between system 1 and system 2, divided by the number of water molecules.